\documentclass[mathleft]{emulatean}
\usepackage{graphicx}
\usepackage{natbib}
\usepackage{url}
\usepackage{times}
\bibpunct{(}{)}{;}{a}{}{}
\usepackage{mathtext,amssymb}

\newcommand{\magdot}[1]{\ensuremath{^{\rm m}\!\!#1\,}}

\newcommand{\AAA}{\ensuremath{\mathrm{\AA}}} %% << added \rm in
\newcommand{\ergl}{~\ensuremath{\rm erg~s^{-1}\/ }}
\newcommand{\ergf}{~\ensuremath{\rm erg~cm^{-2}~s^{-1} \/ }}
\newcommand{\kms}{~\ensuremath{\rm km~s^{-1} \/}}
\newcommand{\keV}{~\ensuremath{\rm keV \/}}
\newcommand{\oiii}{[\ion{O}{III}]}
\newcommand{\oii}{[\ion{O}{II}]}
\newcommand{\oi}{[\ion{O}{I}]}
\newcommand{\sii}{[\ion{S}{II}]}
\newcommand{\nii}{[\ion{N}{II}]}
\renewcommand{\ni}{[\ion{N}{I}]}
\newcommand{\neiii}{[\ion{Ne}{III}]}
\newcommand{\ariii}{[\ion{Ar}{III}]}
\newcommand{\heii}{\ion{He}{II}}

\newcommand{\pc}{~\ensuremath{\rm  pc \/}}
\newcommand{\kpc}{~\ensuremath{\rm  kpc \/}}

\newcommand{\Gyr}{~\ensuremath{\rm  Gyr \/}}
\newcommand{\yr}{~\ensuremath{\rm  yr \/}}
\newcommand{\cmc}{~\ensuremath{\rm  cm^{-3} \/}}

\begin{document}
\Pagespan{0}{}% Document's page range. 
% If second parameter is left empty, the last page is computed automatically.
\Yearpublication{2010}%
\Yearsubmission{2009}%
\Month{}%   
\Volume{}%  
\Issue{}% 
 \DOI{This.is/not.aDOI}% 

\title{The Eastern Filament of W50}
\author{P. Abolmasov\inst{1}\fnmsep\thanks{Corresponding author:
    \email{pavel\_abolmasov@yahoo.co.uk}} \and O. Maryeva\inst{2} \and
   A. N. Burenkov\inst{2}}
\titlerunning{The Eastern Filament of W50}
\authorrunning{P. Abolmasov et al.}
% \begin{document}

\institute{Sternberg Astronomical Institute, Moscow State University,
  Moscow, Russia 119992
\and
Special Astrophysical Observatory,
    Nizhnij Arkhyz,
    Zelenchukskij region,
    Karachai-Cirkassian Republic,
    Russia 369167}

\received{}
\accepted{}
\publonline{}


\label{firstpage}

\abstract{We present new spectral (FPI and long-slit) 
data on the Eastern optical filament of the
well known radionebula W50 associated with SS433. We find  that on
sub-parsec scales different emission lines are emitted by different
regions with evidently different physical conditions.
Kinematical properties of the ionized gas show evidence for 
moderately high ($V \sim 100\,\kms$) supersonic motions. 
\oiii$\lambda$5007 emission is found to be multi-component and differs
from lower-excitation \sii$\lambda$6717 line both in spatial and
kinematical properties. Indirect evidence for very low characteristic
densities of the gas ($n \sim 0.1\cmc$) is found.
We propose radiative (possibly incomplete) shock waves in
low-density, moderately high metallicity gas 
as the most probable candidate for
the power source of the optical filament. Apparent nitrogen
over-abundance is better understood if the location of W50 in the
Galaxy is taken into account.
}

\maketitle

\section{Introduction}

W50 radionebula was first catalogued in the work by \citet{westerhout}
and classified as a radiative-stage supernova remnant by
\citet{holcas69}. However, the nebula is strongly affected by the
activity of the central source SS433 (see \citet{dubner} for review). It is not evident
whether the contribution from the initial supernova explosion, that
probably precursed formation of the compact accretor in SS433
system, may be distinguished from the impact of the jet and wind
activity of the central system. Most of the peculiarities of W50 may
be attributed to the jet activity of SS433. The total power of the
jets is of the order $10^{39}\,\ergl$.
The jets are an immense source of energy
capable to provide the energy of 
$\sim 10^{51}\,\rm erg$ characteristic for a usual
supernova remnant (SNR) in about $3\times 10^4\,\yr$, that is at least
several times lower than the observed dynamical age of W50
\citep{yama}.

The morphology and radio properties of the nebula are fairly
reproduced in numerical simulations with the preceeding SNR
\citep{zavala,velazquez}. However, it is not clear whether the
central, quasi-spherical part of the radionebula requires a preceeding
SNR cavity or not. 
The simulations also do
not consider the possible emission of the warm gas that may appear in
some parts of the nebula where the density is high enough for the
shocked material to cool. 

%%%%%%%%%%%%%%%%%%%%%%%%%%%%%%%%%%%%%%%%%%%%%%%%%%%%%%%%%%%%%%%%%%%%

Warm gas in W50 was detected by \citet{vdberg} as the Eastern
and the Western filaments. Recent investigation by \citep{Boumis}
proved that warm gas emission is present in different parts of the
nebula but peaks in several bright optical filaments offset from the
areas bright in X-ray and radio ranges. 

The W50 optical filaments are distinguished among the known shell-like and
SNR-like nebulae by the strong \nii\ lines that were attributed by
\citet{zealey} to low velocity shocks and by \citet{shuder} to
nitrogen over-abundance.
In this article, we analyse the reasons for the apparent nitrogen
over-abundance, involving Galactic abundance gradients (see \S
\ref{sec:disc:abund} for details).

In the next section we describe the observational material and the methods
used for data reduction. In \S~\ref{sec:res} main results are presented.
The results are discussed in \S~\ref{sec:disc}.

\section{Observations and data reduction}\label{sec:obs}

All the observations were carried out in the prime focus
of the Russian Special Astrophysical Observatory 6m telescope with the
SCORPIO multi-mode focal reducer \citep{scorpio} and EEV~42-40 $2048\times2048$ CCD.
%  Seeing was around $1.5\div 2\arcsec$ during all the observations. 
%The detector was EEV~42-40 $2048\times2048$ CCD.
% During FPI operated with binning
% $4\times 4$ to reduce the readout time. 
%The spatial scale is $0\farcs{}7$ per pixel.

The observational data summary is given in Table~\ref{tab:obslog}.
Exposure times in the table are presented as the duration of a single
exposure multiplied by the number of exposures or (for FPI data) the number of
exposures in a single datacube multiplied by the number of datacubes.

\begin{table*}\centering
\caption{ Observational logs for the long-slit and FPI data. Free
  spectral range is given for FPI data in the ``spectral range''
  column.}\label{tab:obslog}
\begin{tabular}{l@{\hspace{7\tabcolsep}}c@{\hspace{3\tabcolsep}}c@{\hspace{5\tabcolsep}}c@{\hspace{3\tabcolsep}}c}
 & \multicolumn{2}{c}{Long Slit~~~~~~~~~~~~~~} & \multicolumn{2}{c}{FPI}\\
 & Object~~~ & Background  &  \sii$\lambda$6717  &
  \oiii$\lambda$5007 \\
\noalign{\medskip}

Date & \multicolumn{2}{c}{2006 Jul 31~~~~~~~~~~~~~~} &
\multicolumn{2}{c}{2008 Jul 23} \\
Disperser & \multicolumn{2}{c}{VPHG550G~~~~~~~~~~~~} & \multicolumn{2}{c}{FPI501} \\
Spectral Resolution $\delta v$, \kms & \multicolumn{2}{c}{500$\div$1000~~~~~~~~~}  & 31 & 36 \\
Spectral Range, \AAA\ & \multicolumn{2}{c}{3600$\div$7300~~~~~~~~~~~~} & 6717$\pm$7 & 5007$\pm$4 \\
Seeing, \arcsec\  & \multicolumn{2}{c}{1.7~~~~~~~~~~~~} & \multicolumn{2}{c}{1.5$\div$2.2} \\
Exposure Time, s &  900$\times$4 & 600 & 120$\times$36$\times$2 & 200$\times$36 \\
Position Angle, $\deg$ & \multicolumn{2}{c}{37~~~~~~~~~~~~} & 128 and -20  &  -20 \\
\end{tabular}
\normalsize
\end{table*}

\subsection{Long-slit data}\label{sec:LS}

We use archival long-slit spectra of 
the brightest part of the Eastern filament. The slit was centered
on the coordinates $\alpha = 19^h 14^m 18^s$,
$\delta = +5^\circ  03^\prime 20\arcsec{\,}$ (J2000).
This position corresponds roughly to the center of the brightest part
of the Eastern optical filament of W50 \citep{zealey} and is offset
from the X-ray bright lens-shaped region \citep{brinkmann} by about
5$^\prime$ to the North. 
The length of the slit is about 5 arcminutes that is
comparable to the length of the filament itself. This makes it
difficult to subtract the background spectrum, therefore we use night
sky exposures obtained the same night near the spectral standard star
LDS~749B.

The position angle allows to trace the brightest regions of the
filament, better visible in low excitation lines like
\sii$\lambda$6717. 

\subsection{Long-slit data reduction}\label{sec:lsred}

Long slit data were reduced using IDL-based software. The reduction
process includes all the standard reduction steps. Additionally, error
frames are calculated in the assumption that the statistics is
Poissonian for the quanta detected by the CCD. Spectral standard star
LDS749B observed the same night in a 1$^\prime$ annular aperture is a
standard from the list given by \citet{turnshek90}. $4\times 4$
binning was used to reduce the readout noise. 
The reduction process for the night sky exposure was identical
to that used for the object exposures. 
We also used an auxiliary image in the V band with the same
coordinates and position angle to determine the
coordinates of the field stars (see appendix \ref{app:stars} for details).

A significant complication for obtaining high a quality nebular
spectrum is in the large number of field stars. We extract the spectra
of 20 stars roughly brighter than 21$\magdot{\,}$ in V. Stellar spectra
subtraction procedure is described in appendix~\ref{app:stars}.
After subtracting the
contribution from the field stars, the two-dimensional
nebular spectrum still contains some contribution from non-resolved
stellar background. In addition to that, the stellar spectra were
catalogued together with the approximate (accuracy about
1$\arcsec{\,}$) coordinates. We briefly discuss the field star spectra
in appendix \ref{app:stars}. 

Emission line profiles were fitted by gaussian and multi-gaussian
models. The flux ratios of
\nii$\lambda\lambda$6548,6583 and \oiii$\lambda\lambda$4959,5007
doublets were fixed. We also assume the line-of-sight 
velocities and the widths of doublet components are equal. 
Kinematical information is difficult to study using long-slit
spectra having low spectral resolution. However, it is possible to
exclude emission line broadening by hundreds of \kms\ or more. 

\subsection{Scanning FPI data}\label{sec:FPI}

We used a scanning Fabry-P\'erot Interferometer (FPI) providing
spectral resolution  $30\div35$\kms\ to study the kinematics of the
same region. FPI field has a field of view close  to 5$^\prime$ in
diameter and was centered at $\alpha = 19^h 14^m 25^s$,
$\delta = +5^\circ  03^\prime 12\arcsec{\,}$ (J2000). 
The object was observed in two emission lines:  $\sii\lambda$6717 (total
exposure $120\rm\,s \times 36$ spectral channels for each of the two
datacubes) and  $\oiii\lambda$5007 (total
exposure $200{\rm\,s} \times 36$ spectral channels). The two datacubes
obtained for the \sii\ line differ in position angle that allows
to eliminate ghosts from both the bright stars and the nebula
itself \citep{ifpred2}. In the case of the \oiii\ datacube we also use
lower quality data obtained during the next night, July 24, to
remove the ghosts.
The free spectral range was $13.7$ and $7.7$\AA\ for $\sii\lambda$6717
and $\oiii\lambda$5007, correspondingly. 

Data reduction was performed in IDL environment. 
FPI data were reduced using  {\tt ifpwid} software 
designed by Alexei Moiseev. Data reduction algorithms 
are described by \citet{ifpred} and \citet{ifpred2}.
Line profile parameters were determined by fitting with Voigt
functions of fixed Lorentzian widths
($31$\kms\ for \oiii$\lambda$5007 and $36$\kms\ for \sii$\lambda$6717). 
Instrumental profile was measured using the spectra of
a He-Ne-Ar calibration lamp.
Voigt fitting procedure allows to measure line widths even when they
are less than the instrumental profile width \citep{ifpred2}. 
Profiles were fitted only in the pixels where flux
exceeded 20~ADU (corresponding to $S/N \sim 3$ for a given spatial
sampling element).
All the line-of-sight velocities presented here are heliocentric.

Several relatively bright stars were used for relative astrometry
between the datacubes. Accuracy is at least better than the actual
seeing (about $2\arcsec{\,}$).

\section{Results}\label{sec:res}

\subsection{Integral spectrum}\label{sec:intsp}

\begin{figure*}
 \centering
  \includegraphics[width=0.8\textwidth]{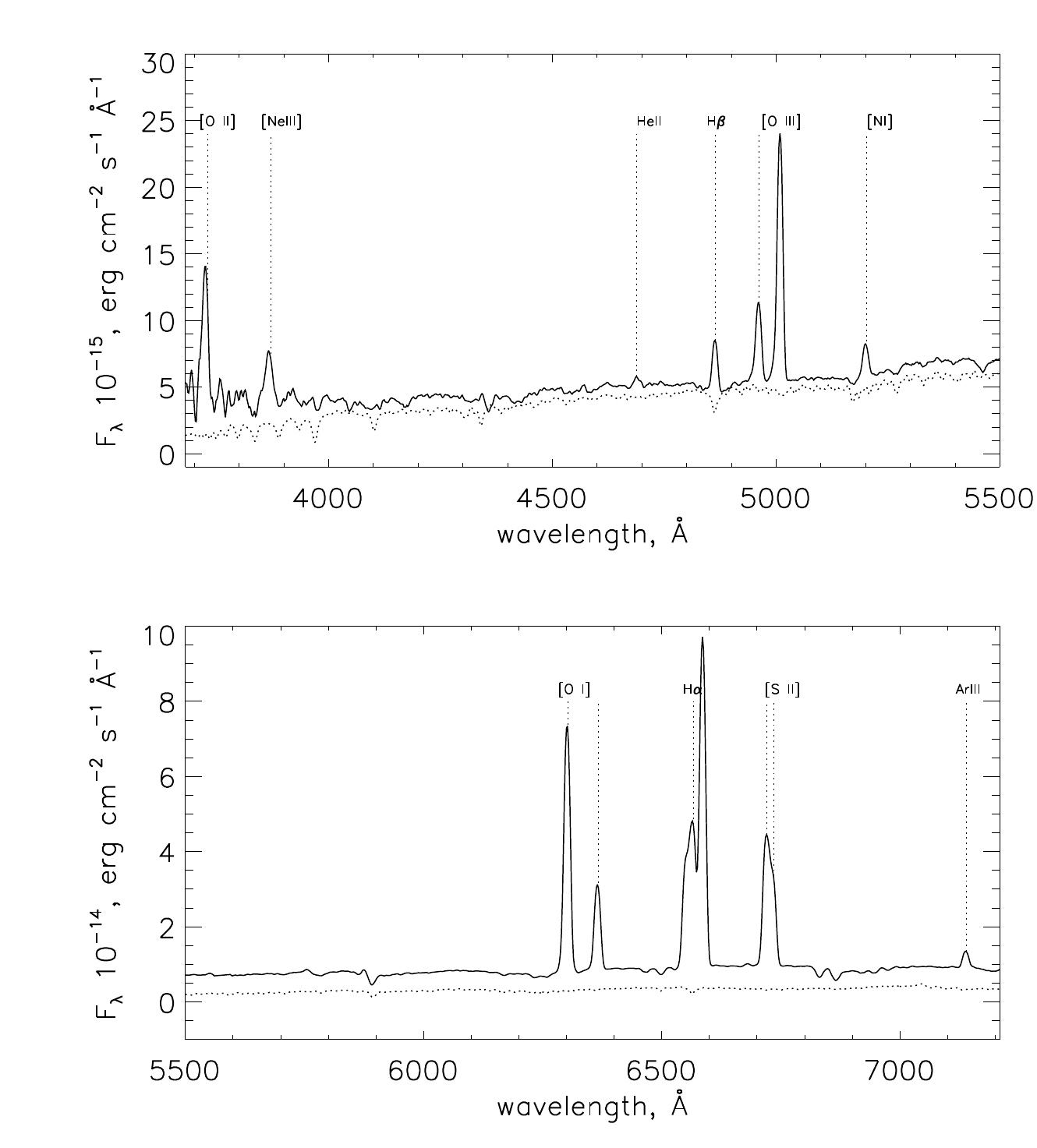}
\caption{The integral spectrum of W50 extracted from the long-slit
  data. Stellar population model spectrum shifted by 10$^{-14}\ergf$
  down is shown for comparison (see
  text for details) by a dotted line.
% Interstellar absorption
%  determined by H$\alpha$/H$\beta$ ratio is $A_{v}$
%  =3\magdot{.}98$\pm$0.21.
}
\label{fig:intsp}
\end{figure*}

The integral spectrum obtained by simple integration along the $\sim
5^\prime$ slit is shown in Fig.~\ref{fig:intsp}. Night sky background
was subtracted using a free
multiplication factor (to account approximately for possible night sky
lines variability) adjusted to accurately zero the
\oi$\lambda$5577 night sky emission in the residual spectrum.

However, the SS433+W50 system is located at a very low Galactic
latitude ($b \simeq -2^\circ$). 
Even after removing 20 field stars and night sky emission
line spectrum there is still a contribution from the Galactic
unresolved background. For comparison, in Fig.~\ref{fig:intsp} we
show a stellar population model from \citet{bruzual} with $T =
12\,\Gyr$ and 1.6$Z_\odot$ metallicity, reddened by $A_\mathrm{V} =
2\magdot{\,}$ (reddening curve by \citet{CCM} with $R_\mathrm{V}=3.1$). This interstellar absorption value was estimated by
$\chi^2$-minimization technique for a fixed population age and
metallicity, the latter chosen in consistence with the nebula position
in the Galaxy (see \S~\ref{sec:disc:abund}). 
% with \citet{chabrier} mass
%function. 
The low interstellar absorption value (compared to the $\sim
4\magdot{\,}$ value given below) may be due to the large number of unresolved
stars in front of the filament
% This is consistent with the moderately
%low interstellar absorption values for the field stars 
 (see appendix \ref{app:stars}).

\begin{table*}\centering
\caption{ Optical emission line spectrum integrated over the long
  slit. }\label{tab:lines}
\begin{tabular}{l@{\hspace{4\tabcolsep}}cc@{\hspace{4\tabcolsep}}cc}
Line &  F, 10$^{-14}\ergf$ & F/F(H$\beta$) &  F, 10$^{-13}\ergf$ & F/F(H$\beta$) \\
 & & &  \multicolumn{2}{c}{( Unreddened by $4\magdot{\,}$ )} \\

   \oii$\lambda$3727 &   15$\pm$1  & 3.16$\pm$0.26 &   360$\pm$30  & 10.83$\pm$0.91 \\
 
   \neiii$\lambda$3869 &   6.2$\pm$0.6  & 1.35$\pm$0.12 &   130$\pm$10  & 3.96$\pm$0.42 \\

   \heii$\lambda$4686 &   0.8$\pm$0.11  & 0.18$\pm$0.02 &   7.2$\pm$1.0  & 0.21$\pm$0.03 \\

  H$\beta$ &   4.6$\pm$0.3  & 1.00$\pm$0.06 &   34$\pm$2  & 1.00$\pm$0.06 \\
 
   \oiii$\lambda$4959 &   9.6$\pm$0.1  & 2.09$\pm$0.02 &    59.3$\pm$0.7  & 1.77$\pm$0.02\\
 
   \oiii$\lambda$5007 &   28.8$\pm$0.3  & 6.26$\pm$0.07 &   178$\pm$2  & 5.30$\pm$0.06 \\
 
   \ni$\lambda$5200 &   3.9$\pm$0.2  & 0.84$\pm$0.04 &   19.3$\pm$0.9  &   0.58$\pm$0.03\\
 
   \nii$\lambda$5755 &   1.7$\pm$0.2 & 0.38$\pm$0.05 &   5.1$\pm$0.6  & 0.15$\pm$0.02\\
 
   \oi$\lambda$6300 &   107.6$\pm$1.0  & 23.42$\pm$0.21 &   210$\pm$2  & 6.27$\pm$0.05\\
 
   \oi$\lambda$6363 &   35.9$\pm$0.3  & 7.81$\pm$0.07 &   70.1$\pm$0.6 & 2.09$\pm$0.02\\
 
   \nii$\lambda$6548 &   45$\pm$5  & 9.81$\pm$1.16 &  73$\pm$9  &  2.18$\pm$0.26 \\
 
   H$\alpha$ &   58$\pm$5  & 12.63$\pm$1.19 &   96$\pm$9  & 2.86$\pm$0.27\\
 
   \nii$\lambda$6583 &   135$\pm$6  & 29.42$\pm$1.33 &   219$\pm$10  &   6.53$\pm$0.29 \\
 
   \sii$\lambda$6717 &   55$\pm$3  & 11.87$\pm$0.61 &   81$\pm$4  &  2.41$\pm$0.13\\
 
   \sii$\lambda$6731 &   36$\pm$3  & 7.76$\pm$0.60 &    53$\pm$4  &   1.58$\pm$0.13\\
 
   \ariii$\lambda$7135 &   7.4$\pm$0.2  & 1.60$\pm$0.05 &   8.5$\pm$0.3  &  0.25$\pm$0.01\\
\end{tabular}

\normalsize
\end{table*}

Integral line fluxes are given in Table~\ref{tab:lines}. Several
emissions are detected for the first time. The most interesing among
these is the \heii$\lambda$4686 line. Though its profile may be
affected by the stellar population spectrum, the line is clearly
identified. The measured \heii$\lambda$4686 / H$\beta$ intensity ratio is as high
as 0.2 that is likely to be the consequence of high electronic
temperature. Line profile may be however affected by the stellar
background, therefore the  uncertainty of its flux may be larger than the
proposed 15\% statistical uncertainty given in Table~\ref{tab:lines}.

Assuming that the temperature of the warm gas is $T_e
\simeq 15\,000\,\rm K$, and the density is lower than $\sim$10\,\cmc, one may
estimate the intrinsic H$\alpha$ /  H$\beta$ intensity ratio as 2.85
\citep{osterbrock}.
Interstellar absorption may be therefore estimated (applying reddening
curve by \citet{CCM}) by the following
formula:

\begin{equation}\label{E:abs}
A_\mathrm{V} \simeq 0.15 \lg \left( \frac{F(H\alpha)/F(H\beta)}{2.85} \right)
\end{equation}

Interstellar absorption is affected by the still poorly known
contribution from the stellar background. If no H$\beta$ absorpion is
present, $A_\mathrm{V}  =4\magdot{.}0\pm0.2$. If the best-fit stellar population model
(see above) is subtracted, $A_\mathrm{V}  \simeq
3\magdot{.}5$. %% 2010 text:
Equivalent widths of higher-order absorption lines H$\gamma$
and H$\delta$ are definitely over-estimated by the stellar population
model (see figure \ref{fig:intsp}), maybe because brighter stars of
intermediate spectral classes were cleaned from the spectrum. The
systematic absorption shift produced by the unresolved fore- and
background stars is therefore no more than $\sim 0\magdot{.}5$ that is
comparable to the absorption variation along the slit (see section
\ref{sec:avgrad}). For the lines at the blue end of the spectrum
(namely, for \oii$\lambda$3727) it introduces an $\sim 2$ uncertainty.
%%%%%%

We estimate the characteristic intensity ratios of \sii\ and \nii\
lines as I(\sii$\lambda$6717) / I(\sii$\lambda$6731) = 1.52$\pm$0.22 and 
(I(\nii$\lambda$6583) + I(\nii$\lambda$6548) / I(\sii$\lambda$5755) =
58$\pm$15, correspondingly. Applying TEMDEN internet service
(\url{http://stsdas.stsci.edu/nebular/temden.html}, see \citet{SD94}
for calculation algorythm description)
allows to estimate the electron temperature as $T_e = 13\,000 \pm 2\,000
\,\rm K$. Electron density is too low to measure it
via \sii\ lines. Sulfur lines give only an upper limit, $n_\mathrm{e} \lesssim
100\,\cmc$. Note that these diagnostic lines are emitted in rather
dense regions of the nebula.
Probably, the real electron density is by two-three orders
of magnitude lower than this, see \S~\ref{sec:disc:cinema}.

\subsection{Kinematics}\label{sec:cinema}

\begin{figure*}
 \centering
  \includegraphics[bb=0 0 350
    300,clip,width=\textwidth]{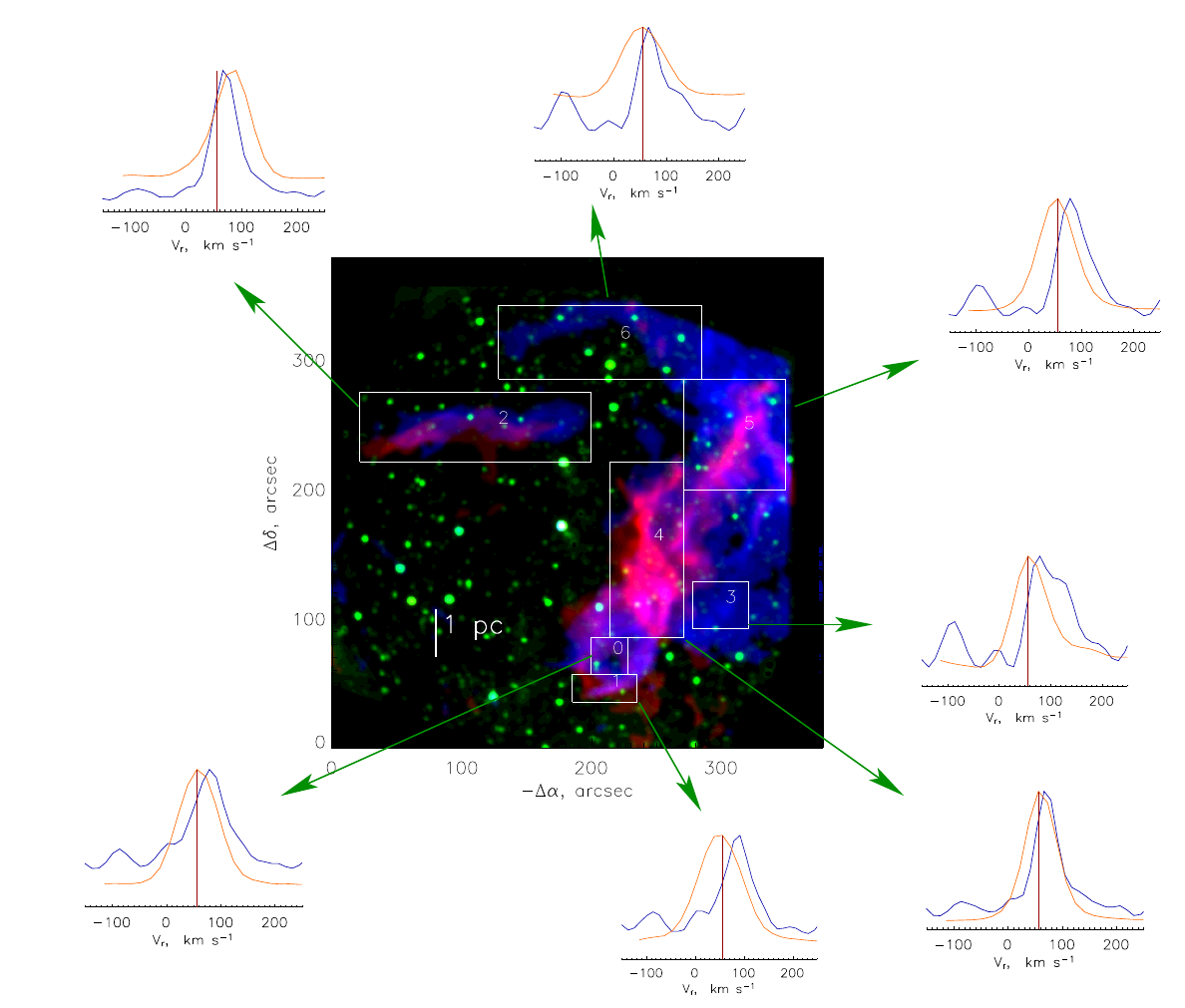} % {profiles6_bw}
\caption{ The two intensity maps overlapped. \sii$\lambda$6717
  intensity is shown by red,
  \oiii$\lambda$5007 by blue (grayscale and dotted contours in the
  black\&white version, dotted lines are logarythmically 
spaced by a factor of 10). 
For seven regions of
  interest line profiles are also given (red/dotted for \sii\ and
  blue/solid for \oiii). The vertical line everywhere has the velocity
  of 56\kms.}
\label{fig:profiles6}
\end{figure*}

Kinematical properties, as well as the morphology of the filament, differ for the \sii\ and
\oiii\ emissions. The fine subparsec-scale
structure of the filament (that is still much coarser than the actual
seeing) is present for the low-ionisation sulfur line but not for
\oiii$\lambda$5007. Most probably, the \sii$\lambda$6717 
line is emitted by denser inner
parts of the filamentary structure while the outer, hotter and more
rarefied layers are better visible in higher-ionization lines. The
picture is similar to the shocked cloud structure in radiative-stage
SNRs. We show the two intensity maps and several selected line
profiles in Fig.~\ref{fig:profiles6}. The 56\,\kms\ mark corresponds
to the line-of-sight velocity reported by \citet{Boumis} as the systemic
velocity.

\begin{figure*}
 \centering
  \includegraphics[width=0.75\textwidth]{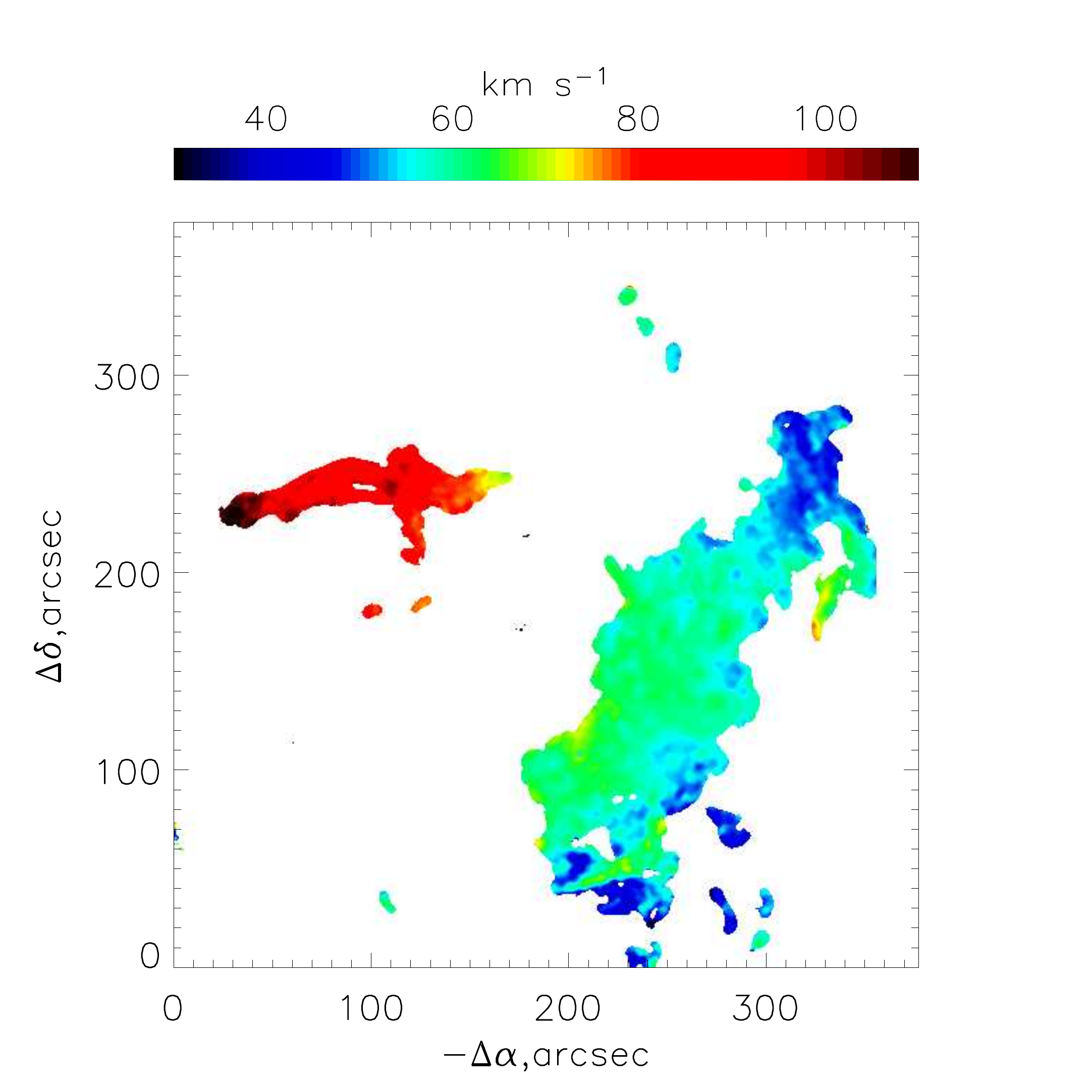} % {s2_v} 
\caption{Velocity map for \sii$\lambda$6717.}
\label{fig:s2_v} 
\end{figure*}

The velocity map for \sii$\lambda$6717 is given in
Fig.~\ref{fig:s2_v}. The brighter part of the filament shows
practically
constant line-of-sight velocity of 50$\div$70\,\kms\ with about 20\,\kms\
variations near the bubble structure covered by regions 0 and 1 in
Fig.~\ref{fig:profiles6}. The line is broader than the components
of \oiii\ and is generally broadened by 15$\div$20\,\kms. In the
regions 1 and 6 the line has larger velocity dispersion of about 40\,\kms. 
The most outstanding feature is the longitudally oriented smaller
filament (region 2 in Fig.~\ref{fig:profiles6}). It is evidently
kinematically shifted from the main body of the filament. Its
line-of-sight velocity is in the range of 80$\div$100\,\kms.

The profiles of \oiii$\lambda$5007, on the other hand, show more
complicated structure. In most of the regions of interest two to four
components with velocity dispersion about 10\,\kms\ are present: the brighter one at 70$\div$80\,\kms, slightly shifted with
respect to the sulfur line, one at 120$\div$130\,\kms\ and
blueward-shifted components at -10$\div$0 and -80$\div$-100\,\kms. 

\section{Discussion}\label{sec:disc}

\subsection{Abundances}\label{sec:disc:abund}

The crucial point in understanding the emission line spectrum of the
nebula is in the distance towards the Galactic center. Let us assume
the distance from the Sun is $ 5\div 6\,\kpc$ \citep{lockman}. 
The Galactic latitude and longitude are equal to $b = -2^\circ.3$ and
$l=39^\circ.8$, correspondingly, and the distance from the Sun towards
the Galactic center $7.5\div 8.5\,\kpc$ (see for example
\citet{Eisenhauer} and references therein). We come to the conclusion that
SS433 and W50 are situated at a Galactocentric distance of about
$4\div 5\,\kpc$, i. e., at least $3\,\kpc$ closer to the center than the
Sun. That suggests the metallicity of the gas should be somewhat
higher than solar. Using the Galactic abundance gradient values given by
\citet{alibes}, one may estimate the oxygen abundance as ${\rm [O / H]}
\simeq 0.1\div 0.2$ (relative to the Solar value). For nitrogen the
enrichment may be as high as ${\rm [N / H]}
\simeq 0.2\div 0.3$ because its abundance scales non-linearly with
metallicity and its gradients are generally steeper at higher
metallicities \citep{nitrozee}. Here we consider ``solar'' abundances
$12+\lg {\rm O/H} = 8.7$ and $12+\lg {\rm N/H} = 7.9$ according to
\citet{abundsolar}.

High ambient metallicity makes the unordinary spectrum of W50 more
reasonable. In a solar metallicity environment, \nii$\lambda$6583 over
H$\alpha$ intensity ratio for a W50 analogue is likely to be $\sim
1$. Together with the oxygen line intensity ratios $\oi\lambda 6300
/ \oiii \lambda 5007 \sim 1$ and $\oii\lambda 3727
/ \oiii \lambda 5007 \sim 2$ it makes the filament a bona fide
shock-powered nebula \citep{baldwin}. The observed $\oiii \lambda
5007 /  {\rm H}\beta \sim 5$ ratio is also quite reasonable, if high
ambient metallicity is taken into account. 

Some high ionization
features like the \heii$\lambda$4686 emission (and still high \nii,
\oiii\ and \neiii\ line intensities) may be attributed to incompleteness
of the shock waves. 
Though cooling time is short enough for the observed shock waves to
radiate most of their energy, they may still remain incomplete (i. e.,
lack lower-temperature gas) if
additional energy sources such as X-ray and EUV radiation from SS433
or X-ray bright regions of the nebula are present. Shock multiplicity
also restricts the sizes of cooling regions. It should be also noted
that the spatial cooling region scale is much larger than the
projected width of the long slit, therefore the nebular regions traced
by the slit may represent gas temperatures in unequal proportions.
% differing
%from those appearing if all the cooling region emission is
%intergated. 
 Due to all these reasons, shocks in W50 may bear
incompleteness signatures even though their cooling times are considerably
shorter than the age of the nebula. 

\subsection{Kinematical structure}\label{sec:disc:cinema}

%Intensity distribution in the two emission lines and line profiles are
%given in Fig.~\ref{fig:profiles6}.
Different kinematical behaviour of the two forbidden emissions may be
explained qualitatively by a system of moderately fast (from tens to
$\sim 100\,\kms$) shock waves. Large-scale turbulent motions in
recently shocked gas are
transformed into microturbulent motions in cooler gas emitting the
\sii$\lambda$6717 line that has significantly lower ionisation
potential. Due to this reason, several velocity components transform
into general line broadening (lower-velocity motions at smaller spatial
scales) for lower ionisation lines. 

Spatial displacement between the regions
emitting in the two lines is mostly of the order arcseconds, up to
$10\div 20\arcsec{\,}$. That gives (if we assume that the spatial
structure reflects gas cooling behind shock fronts)
approximate cooling region width $\sim 0.1\div 1\,\pc$. 
This range may be used to estimate the density of the shocked
gas. If the matter moves with velocity $u$ (presumably the same order
with the shock front velocity), it cools from $T_2$ temperature to
$T_1$ at the lengthscale:

$$
l \simeq \frac{u (T_2 - T_1) }{n_\mathrm{e} \Lambda}
$$

Here $n_\mathrm{e}$ is electron density, and $\Lambda$ is cooling
function. $n_\mathrm{e}$ may be estimated from the observational data, if $l$
and $u$ are known. Cooling function
$\Lambda \sim 10^{-21}\div 10^{-22}\,\rm erg \,cm^{-3} \, s^{-1}$
for $T \sim (1\div 1.5)\times 10^4\,\rm K$, hence:

$$
\begin{array}{l}
n_\mathrm{e} \sim 0.1 \left(\frac{u}{100\,\kms}\right)
\left(\frac{T}{10^4\,\rm K}\right)\left(\frac{l}{0.5\,\pc}\right)^{-1}\times\\
\qquad{} \times \left(\frac{\Lambda}{10^{-22}\,\rm erg \,cm^{-3} \,
  s^{-1}}\right)^{-1} \cmc\\
\end{array}
$$ 

Gas density is therefore likely to be well below the lower limits set
by the \sii\ doublet estimates (see \S \ref{sec:intsp}). 
 The corresponding cooling timescale is $\sim l/u \sim 500\,\yr$ that
 is significantly lower than the proposed age of the nebula.

\bigskip

The Eastern optical filament is powered by
shock waves with $V \sim 100\kms$. This characteristic 
velocity is in good agreement
with the value given by \citet{Boumis} for the expansion velocity of
the nebula. The motions seem to be chaotic and may be the
result of the post-shock turbulence of stronger shock waves that
produce the X-ray emission in X-ray bright regions of W50
\citep{brinkmann}. \citet{brinkmann} do not provide any shock velocity
estimates, but they estimate the thermal component temperature as $T
\sim 0.2\,\keV$, that yields $T_S \sim 400\,\kms$ for Sedov solution
(see \citet{ham83}). Strong radiative shock waves are predicted to
produce supersonic turbulence with average Mach number $\sim
0.2\mathcal{M}$, where $\mathcal{M}$ is the upstream Mach number of
the primary shock \citep{FW08}. For an isothermal primary shock, that means also
a factor of 5 decrease of the characteristic velocity. That
approximately accounts for the observed velocity scales present in
W50 (except for the jet velocity itself) as well as for
the fine structure formation itself.
%Numerical simulations of fast shock waves (such as \citet{WF98}) show
%that a $\sim 1000\,\kms$ radiative shock wave is able to produce
%supersonic turbulent field with characteristic velocity $v_t\sim
%100\,\kms$. 
%Thin shell and filament structures appear as well. 

Otherwise, the moderate power shocks may trace the
 expansion of the gas heated in the bow-shock of the Eastern
jet. Thermal velocity in the gas that emits the observed X-ray
radiation ($T \sim 0.2\,\keV$, see above) 
is of the order $\sim 100\div 200\kms$,
that directly yields the required velocity scale. A
pressure-driven shell solution \citep{castor} 
produced by a power source $L \sim
10^{39}\ergl$ and expanding into ISM with $n \sim 1\,\cmc$ will have
expansion velocity:

$$
\begin{array}{l}
V \simeq 0.6 K^{5/3} \left(\frac{L}{\rho}\right)^{3/5} R^{-2/3} \simeq\\ 
\simeq 100 \left(\frac{L}{10^{39}\ergl}\right)^{1/3}
\left(\frac{n}{1\cmc}\right)^{-1/3} \times \\
\qquad{} \qquad{} \times \left(\frac{R}{50\pc}\right)^{-2/3} \kms \\
\end{array}
$$

Here, $K \simeq 0.76$ is a dimensionless constant. If the expansion of
W50 is fed by the power of the relativistic jets, it is likely to have
characteristic velocities similar to the observed $V\sim
100\,\kms$. However, it is difficult to interpret the observed
velocity components as the two walls of a single shell. Expansion
should proceed in a more complicated way than spherically-symmetric
bubble expansion.

%%%%%%%%%%%%%%%%%%%%%%%%%%%%%%%%%%%%%%%%%%%%%%%%%%%%%%%%%%%%%%%%
% 2010 text:

\subsection{Absorption gradients}\label{sec:avgrad}

SS433 and W50 are not only heavily absorbed, but the absorption itself
is patchy, from about 8$^m$ for the central object towards $\lesssim
4^m$ for some parts of the Eastern filament. Variable absorption is
important for the observed morphology of the filaments on the scales
larger than minutes \citep{Boumis}. 

\begin{figure*}
 \centering
  \includegraphics[width=0.75\textwidth]{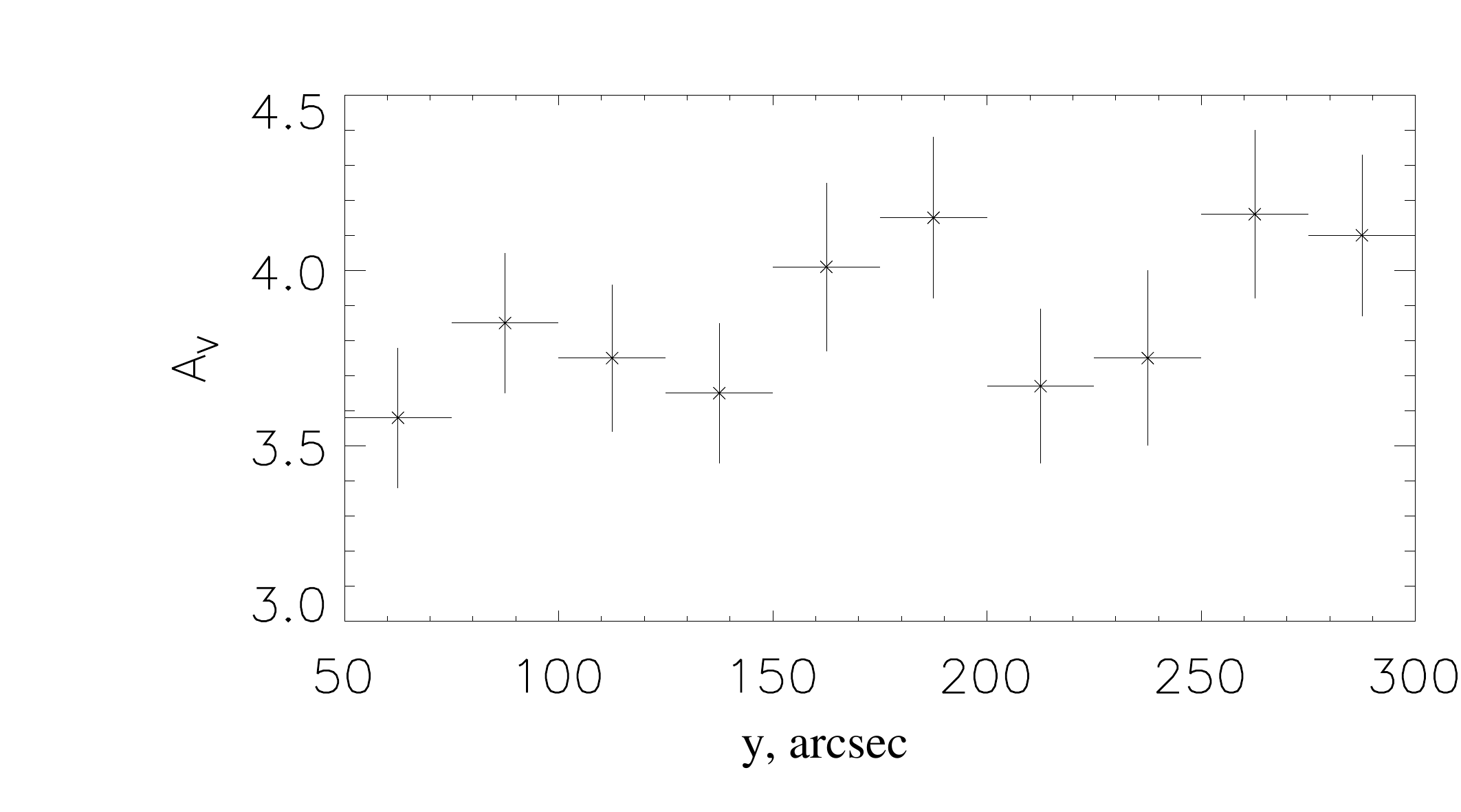} % {s2_v} 
\caption{Spatially-resolved V-band interstellar absorption as function
of coordinate along the slit}
\label{fig:avgrad} 
\end{figure*}

In figure \ref{fig:avgrad} we show the V-band absorption calculated
using the long-slit data devided into 10 bins. For each, we performed
the procedure identical to this made for the integral spectrum in
section \ref{sec:intsp}. Error bars are 1$\sigma$ approximation
uncertainties. $A_V$ changes by about $0\magdot{.}5$, and therefore
may be responsible for about 50\% flux variations for lines near
$5000\AAA$. 
In contrast, emission line intensities  along the slit
vary by about an order of magnitude. 

Variations induced by variable absorption 
hardly affect the morphology in emission
lines. Firstly because the observed fluxes are definitely linked with
kinematics (see \S~\ref{sec:cinema}). Besides this, in regions of
very low \sii\ fluxes we observe bright emission in the \oiii\ line,
that is possible only due to physical reasons. Last but not least, the
bright regions covered by the long slit do not have absorption much
lower than the fainter parts of the filament.

%%%%%%%%%%%%%%%%%%%%%%%%%%%%%%%%%%%%%%%%%%%%%%%%%%%%%%%%%%%%%%%%%%%

\section{Conclusions}\label{sec:conc}

We come to the conclusion that the optical filaments of W50 are mainly
powered by shocks with velocities of the order 100$\div$200\kms\ and less. Flows
with line-of-sight velocities differing by $\pm$100\kms\ are detected
in the \oiii$\lambda$5007 emission line produced in the recently
shocked, relatively hot gas. For lower-ionisation lines kinematics is
different: we detect only general broadening of the
\sii\ line. Probably, cooler gas shows both lower turbulent velocities
and larger number of velocity components. 
We also find at least two spatially resolved shock fronts in the FPI
field of view propagating in different directions. Their line-of-sight
velocities differ by about 50\kms. Our results are compolementary to
the results reported by \citet{Boumis} and provide more detailed
picture of a smaller part of the nebula. 

The optical emission-line spectrum of W50 Eastern filament bears the
main signatures of a radiative (possibly, incomplete) shock 
such as enhanced collisionly excited lines of different ionization
potentials. The observed gas shows a broad range of physical
properties.
In general, the gas is rarefied ($n \sim 10^{-2}\div 1\,\cmc$) and
moderately over-abundant in oxygen and nitrogen. 
%The main reasons for incompleteness are probably the low density
%$n_\mathrm{e} \sim 0.1\,\cmc$ of the shocked gas and additional energy sources
%that complicate cooling of the warm gas.
This apparent over-abundance is likely to be attributed to the high
metallicity of the Galactic gas at the comparatively low ($\sim 4\div 5\,
\kpc$) distance from the Galactic center. Nitrogen lines are primarily
affected because of stronger nitrogen gradient in the Galactic disc. 

\acknowledgements
We thank A.~Moiseev for help with FPI data reduction.

%\newpage

\bibliographystyle{mn2e}   % if natbib is available
\bibliography{mybib}

\newpage

\appendix
%\begin{appendices}

\section{Field Stars Spectroscopy}\label{app:stars}

In order to obtain better integral long-slit spectrum, field stars
were cleaned from the 2D-specta of the object and night sky
background. Stars were identified automatically on
spectrally-integrated one-dimensional images. Stellar profiles were
approximated by Gaussian function (using Moffat function does not
increase $\chi^2$ significantly but decreases stability of the
algorythm) with parameters polynomially dependent on the wavelength. 

20 field stars (roughly brighter than 21$\magdot{\,}$) were finally extracted
from the object frame. To our knowledge, 
none of the stars are present in existing
stellar catalogues. We suggest that the data may be of some use
for future astrophysical applications. We identify the spectra
automatically by $\chi^2$-fitting with spectral standards from STELIB
(\url{http://webast.ast.obs-mip.fr/stelib}, see \citet{stelib}) 
simultaneously estimating interstellar absorption
(\citet{CCM} extinction curves with $R_\mathrm{V} = 3.1$ were used). We
summarize our results in Table~\ref{tab:stars}. 
Spectral classes are given with accuracy about 1$\div$2 spectral
subclasses. Spectral resolution does not allow to determine the
luminosity class self-consistently, but we still give the 
best-fit luminosity classes. Coordinates are measured with accuracy $\sim
    1\arcsec{\,}$ and given in J2000. Coordinates were
measured by comparison of the auxiliary SCORPIO image (see
\S~\ref{sec:lsred}) to the Palomar Digitized Sky Survey 
image\footnote{Based on photographic data of the National Geographic Society -- Palomar
        Observatory Sky Survey (NGS-POSS) obtained using the Oschin Telescope on
        Palomar Mountain.  The NGS-POSS was funded by a grant from the National
        Geographic Society to the California Institute of Technology.  The
        plates were processed into the present compressed digital form with
        their permission.  The Digitized Sky Survey was produced at the Space
        Telescope Science Institute under US Government grant NAG
        W-2166.} 
with correct astrometry. {\tt starast} procedure written by
W.~Landsman was used to set the coordinate grid on the SCORPIO image
and, subsequently, along the slit.

It is easy to check that most of the stars in the observed sector of
the Galaxy and the relevant range of magnitudes
 should be foreground main sequence stars of the spectral
classes from A to K. This is confirmed by the relatively low
interstellar absorption for most of the stars. A $1\arcsec{\,}\times
300\arcsec{\,}$ slit covers about $3\times 10^{-8}$ of the
Galactic volume, that should contain about 1000 main sequence stars,
several white dwarfs and probably no giant and supergiant
stars. If one considers only the nearest parsec of the volume and only
the brightest stars (say, $M_\mathrm{V} \lesssim 5\magdot{\,}$), their number should be
of the order 10$\div$50. 
Most of the stars have visual magnitudes in the
range $18\div 21\magdot{\,}$. We do not however give the best-fit magnitude
values because of the unkown slit losses. 

\begin{table}\caption{Field stars: spectral classes, best-fit
    interstellar absorption and coordinates.}\label{tab:stars}
\center{
\begin{tabular}{cc@{\hspace{1\tabcolsep}}c@{\hspace{3\tabcolsep}}c@{\hspace{1\tabcolsep}}c}
 
 &  Spectral Class  &  $A_{v},\, \magdot{\,}$ & $\alpha$ &  $\delta$\\
 
1 & G7 V               & 0.5     &  19$^h$14$^m$12$^s\!\!$.0 &    +05$^\circ$05$^\prime$08$\arcsec{\,}$ \\
 
2 & K0 V               & 1.5     &  19 14 18.0 &   +05 03 05 \\
 
3 & G9 III               & 2.2   &  19 14 17.2 &    +05 03 22 \\
 
4 & F2 V               & 2.8     & 19 14 22.1   &  +05 01 33   \\
 
5 & G9 III               & 3.2    &  19 14 13.2 & +05 04 39    \\
 
6 & G0 V               & 0.9    & 19 14 18.5    & +05 02 54  \\
 
7 & G8 V               & 3.8    &    19 14 22.4    &  +05 01 31     \\
 
8 & K0 V               & 0(?)   &    19 14 23.7      &   +05 01 07  \\  
  
9 & K1 V               & 0.7   &    19 14 22.0      &    +05 01 41  \\
  
10& K1  III              & 5.6   &    19 14 23.6      &   +05 01 09 \\

11    & K1 V        &  4.3  &   19 14 24.0      &    +05 01 03       \\
 
12    & G8 V         &  3.3  &   19 14 23.0     &    +05 01 23      \\
  
 13   &  G0 V        &  2.7 &     19 14 24.4     & +05 00 49        \\

 14   &  G0 V         &  3.6    &     19 14 19.2 &   +05 02 35       \\

 15   & K2 V         &  2.0    &   19 14 17.2     & +05 03 13        \\
 
 16   &  K4 III      &  3.4 &   19 14 21.6       &  +05 01 41        \\

 17   & K4 III       & 1.7   &    19 14 15.9      & +05 03 45      \\

 18  & G9 III        & 1.2  &   19 14 15.2      &   +05 04 00      \\
  
 19  & G8 V        & 6.0  &   19 14 24.2      &   +05 00 56      \\
    
 20  & A9 III        & 3.9  &   19 14 19.1      &   +05 02 43      \\
      
\end{tabular}
}
\end{table}

% \section{Field Stars}\label{app:fstars}

% \end{appendices}

%\begin{thebibliography}
%\end{thebibliography}

% \end{document}

\label{lastpage}

\end{document}